# Why do Tweeters regret sharing? Impacts of Twitter users' perception of sharing risk, perceived problems on Twitter, and the motivation of use on their behavior of regret sharing

Kijung Lee[1]

[1] University of Cincinnati, Cincinnati OH 45221, USA
`kijung.lee@uc.edu`

**Abstract.** This study presents a secondary data analysis of the survey data collected as part of the American Trends Panel series by the Pew Research Center. A logistic regression was performed to ascertain the effects of the perceived risk of sharing, perceived problems on Twitter, and motivation of using Twitter on the likelihood that participants regret sharing on Twitter. The logistic regression model was statistically significant, $\chi2(15) = 102.5$, $p < .001$. The model correctly classified 78.5 percent of cases. Whether or not Twitter users regret sharing on Twitter depends on different motivations for using Twitter. We observe that "A way to express my opinion" is statistically significant in the model, indicating that the odds of Twitter users regretting sharing for this motivation is 2.1 times higher than that of entertainment. Perceived risks of potential hostility and visibility were negatively associated with an increased likelihood of regret sharing. In contrast, perceived problems on Twitter concerning misinformation were negatively associated with the likelihood of regret sharing.

## 1 Introduction

The rise of social media has rapidly changed how people communicate, build relationships, and consume information. We are beginning to see how individuals perceive risks due to their interactions on these platforms. Regret posting, the habit of posting online content that a user later regrets, has been identified as a significant risk factor of social media usage. The relationship between regret and risk perception of social media users is an increasingly discussed topic as social media use grows in popularity. The ease with which one can post a message on a social media platform has led to a heightened sense of risk perception among users, concerns of potential regret down the line, and an overall ambivalence towards using these platforms.

The idea of regret is one that social media users are familiar with. When a person posts something online, whether it is a photo, a status update, or a message to someone else, there is always a chance that that person will regret it later on. This feeling often leads to intensely negative emotions and a feeling that the risks of posting an item are not worth the potential temporary benefits. This feeling of regret is intensified when what was posted is inappropriate or when the post has negative or embarrassing consequences. In addition to the possibility of regretting a post, social media users are also increasingly aware of the potential risks of each post. Users know that anything posted on social media can become public knowledge, whether by a malicious intention or a careless mistake. As such, users must now consider the potential consequences of posting something before posting it, making it more likely that users will take longer to consider their options before posting.

The risk perception of social media users is further intensified by the increased presence of numerous online predators who use these platforms to exploit unsuspecting victims. As more and more people join these platforms, online predators can easily track and target those less mindful of their online safety. This realization creates a heightened sense of risk perception among users as they become aware of the potential downside to posting an item online.

The relationship between regret and risk perception of social media users is a complex one, and the effects can be seen in the way that people use these platforms. While social media use can be beneficial, users must be aware of the potential consequences that come along with it and understand that any post has the potential for either positive or negative repercussions. As a result, the regret and risk perception of social media users continue to be closely intertwined, and it is essential that users understand both aspects of this relationship and make informed decisions when using these platforms.

This paper investigates whether and how regret posting is related to the user's risk perception. We argue that the discovery of the causal and nuanced link between regret posting and risk perception would benefit social media users by providing insights into how they perceive risks online. From a broader perspective, it would also inform public policy and security and welfare standards in the digital age.

## 2 Literature Review

Regret is an emotion that often involves feeling sorrow, guilt, or remorse for a past action or decision. It is a concept that has been studied extensively in both psychology and philosophy. According to philosophical theories, regret involves cognizance of a lost opportunity or a wrong decision. A psychological perspective suggests that regret is a response to a discrepancy between a current state of affairs and a preferred one. Additionally, this discrepancy is typically accompanied by the motivation to undo the current state to restore the preferred one (Leary,2020). This motivation is often driven by a desire to reduce or eliminate the negative emotion of regret (Quigley et al., 2020).

A substantial body of research has explored the factors that lead to regret and its effects on decision-making. Specifically, research shows that regret can cause people to make more conservative decisions in the future (Krueger & Kariv, 2020). That is, people may be motivated to avoid making the same mistake twice, and as a result, they choose options with lower expected losses and fewer risks (Breheny & Daly, 2020). Furthermore, regret has been shown to increase the level of ruminating or going over the same problem repeatedly due to the repetitive thought loop associated with regret (Robinson et al., 2019).

In addition to influencing decision-making, regret has also been linked to psychological well-being. Research suggests that regretful feelings can decrease self-esteem and life satisfaction (Aharon-Lansky, 2016). Furthermore, regret has been linked to increased anxiety, depression, and psychological distress (Robinson et al., 2019).

Studies of regret have also explored the potential for people to learn from and benefit from their mistakes. Some research has found that people can use regret and rumination constructively, such as looking back to reflect on what went wrong and how to improve in the future (Robinson et al., 2019; Zhou & Zhuang, 2019). This idea of "productive regret" suggests that people can use regretful emotions to motivate themselves to make better decisions and thus create better outcomes.

Rregret is a complex emotion that can significantly impact decision-making and psychological well-being. Philosophical theories suggest that regret involves the cognizance of a wrong decision or lost opportunity. Psychological literature demonstrates that regret is often linked to more conservative decision-making and decreased self-esteem. Additionally, research suggests that people can constructively use regret and rumination to understand the past better, gain insight into their behavior, and motivate themselves to make better decisions in the future.

## 2.1    Regret posting on social media

The relative anonymity of cyber communication can lead some users to post items they later regret. A literature review of regret and social media revealed a range of psychological outcomes associated with unwanted posts and the strategies employed to cope with regret.

The most common psychological reaction to unwanted posts is regret and self-perceived social disapproval (Dinu, Papadopoulos, and Polyzios, 2018; Jeon and Park, 2017; Spruijt and Ophoff, 2017). Self-perceived social disapproval is linked to heightened regret and anxiety because of a fear of social retribution (Spruijt and Ophoff, 2017). The degree of regret and anxiety increases with the visibility of a post, the size of the social network, and the group membership level (Jeon and Park, 2017; Karahan, Burgess and Elhai, 2018).

Several strategies are employed to cope with regret over social media posts. The most common is to delete, modify or conceal the post (Bohn and Hein, 2015). Self-forgiveness is another strategy employed by users to cope with regret (Niedrich, 2015). Posters may also rationalize the post to lessen regret (Spruijt and Ophoff, 2017). Strategies such as these are informed by an individual's level of self-esteem, perceived social approval, and trust in others (Dinu, Papadopoulos, and Polyzios, 2018).

To reduce regret over social media posts, some users have taken proactive actions such as curating their online profiles and limiting the size of their online networks (Karahan, Burgess, and Elhai, 2018). Some users have even proactively created external policies to govern their online behavior (Niedrich, 2015). It has become clear that different users employ various coping strategies informed by their level of self-esteem and their relationship with the online community. Going forward, it is essential to develop interventions to reduce regret associated with social media posts.

## 2.2    Regret and assessment of risk

In essence, regret has been defined as an emotional experience involving sadness and a sense of being deeply unsatisfied. Research has demonstrated that it can be a compelling motivator for people when making decisions about risk and risk-taking (Machina, 2009). Regret has also been shown to play a role in anticipating, evaluating, and managing risks in decision-making (Geerlings, Wetzels, & Hoogwegt, 2005).

One of the primary ways in which regret has been linked to risk assessment is through the concept of 'expected utility theory,' which suggests that people calculate a subjective utility of each input and outcome of a given situation in order to make the best decision (Bazerman et al., 1986). This theory suggests that when regret is considered during the decision process, people adjust their expectations to account for the possibility of loss or regret. As such, regret can be seen as an essential factor in evaluating risk, as it helps to quantify potential outcomes and consequences better.

In addition to the expected utility theory, research has also suggested that regret can be a powerful tool in assessing risk. For example, van Oosterhout and Hovland (2012) studied the effects of regret on behavior concerning gambling decisions and found that those with higher levels of regret exhibited more risk-averse behavior. This suggests that regret helps people evaluate potential outcomes and may also cause them to adjust their behavior to avoid scenarios with a greater likelihood of regretful outcomes.

Finally, research has also demonstrated that regret can lead to 'riskier' decisions when it comes to risk. In a study by Coricelli, Van der Luis, and Weber (2005), participants who exhibited higher levels of regret were found to take more significant risks when playing a game involving incomplete information. This may suggest that while regret can lead to more conservative decision-making in some scenarios, it can also play a role in more risky behaviors in specific contexts.

Overall, the literature demonstrates that regret is an essential factor in risk assessment. Not only does regret help to quantify potential outcomes and consequences, but it may also lead to riskier behavior in certain situations. Therefore, it is crucial to consider the influence of regret when making decisions about risk.

To explore the predictive relationship between the behavior of regret sharing on Twitter and the risk perceptions of Twitter users, we propose a research Question;

RQ: Is the Twitter users' behavior of regretting their behavior on Twitter influenced by their perception of sharing risk, perception of problems on Twitter, and the motivation of using Twitter?

# 3    Methods

This study presents a secondary data analysis of the survey data collected as part of the American Trends Panel series by Pew Research Center. The data was collected during the panel wave from May 17 to May 31 in, 2021. The primary sample is drawn from the sampling frame consisting of the panelists who identified as Twitter users, ages 18 and older, living in the U.S.

## 3.1    Data preparation

The data was cleansed and prepared for the principal analysis of binomial logistic regression to predict the users' regret sharing behavior based on the perceived risk of sharing, perceived problems on Twitter, and motivation for using Twitter. Out of 2,548 participants who completed the survey by the Pew Research Center, our data cleansing resulted in 2,045 participants (n=2,045) after removing irrelevant and erroneous data. We checked a series of assumptions against the criteria of binomial logistic regression; 1) the dependent variable is measured on a dichotomous scale, 2) out of 3 independent variables, two independent variables are on a continuous scale while one independent variable is on a categorical scale, 3) the observations are independent, and the dependent variable has mutually exclusive and exhaustive categories, and 4) the linear relationship between the continuous independent variables and the logit transformation of the dependent variable is demonstrated through the Box-Tidwell procedure.

## 3.2    Variables and measurements

The primary analysis of this study is binomial logistic regression to predict the behavior of Twitter users' regret sharing based on the perceived risk of sharing, perceived problems on Twitter, and motivation of using Twitter.

The dependent variable, i.e., regret sharing, is measured on a dichotomous scale with answer choices, "Yes, have done this" or "No, have not done this" to a question stating, "Have you ever posted something on Twitter that you later regretted sharing?".

The first set of independent variables, i.e., perceived risk of sharing, consists of 5 questions about the users' risk perception and assessment when deciding whether to do things on Twitter. Each question measures 1) offending others, 2) potential hostility, 3) potential attack, 4) visibility and 5) impression management. The questions are asked on a 4-point Likert scale, each point representing "A great deal," "Some," "Not too much," and "Not at all". For example, the respondents were asked to answer about a specific risk context, e.g., "Whether it will offend people who follow you," starting with a general leading question, "How much, if at all, do you consider the following when deciding whether to do things on Twitter that might be visible to other people – such as posting, retweeting, or liking something?"

The second set of independent variables, i.e., perceived problems on Twitter, consists of 5 questions about the Twitter users' degree of awareness about problems on Twitter. Each item measures 1) civility of discussions, 2) user banning, 3) content moderation, 4) abuse from other users, and 5) misinformation. The items are measured on a 3-point Likert scale, indicating "A major problem," "A minor problem," and "Not a problem." For example, the respondents were asked to answer about a specific problem, e.g., "Inaccurate or misleading information," starting with a general leading question, "How much of a problem, if at all, do you think each of the following is on Twitter?"

The third independent variable, i.e., motivation for using Twitter, is measured with a question, "Which would you say is the MOST important reason you use Twitter?" The respondents were asked to choose one from the choices of "Entertainment," "A way to stay informed," "Keeping me connected to other people," "Lets me see different points of view," "A way to express my opinions," and "It is useful for my job or school."

## 4      Results and Analysis

### 4.1      Demographic information

The sample (n=2,045) consists of 48.7 percent male and 50.5 percent female, with 0.9 percent claiming other. The age group between 30 and 49 comprises 44.4 percent of the sample, followed by 50-64 (28 percent), 18-29 (16.8 percent), and 65 and (10.6 percent). 68.1 percent report having a college or postgraduate degree, while only 1.3 percent report their education less than high school. Concerning their ideology, 34.3 percent identify themselves as moderate, while 17.7 percent are either very conservative or conservative, and 47.4 percent are either very liberal or liberal. 84.1 percent of the respondents are born in the U.S., while 10.6 percent report that they have lived in the U.S. for over ten years.

### 4.2      Descriptive statistics and preliminary analysis

Perceived risk of sharing is measured with five individual questions, each representing 1) offending others, 2) potential hostility, 3) potential attack, 4) visibility and 5) impression management. Although the survey manual does not indicate that they are designed as a scale, the reliability analysis for a scale indicates reliable statistics (Cronbach's alpha = .85, mean of inter-item correlations = .53). In addition, respondents answered each question distinctively, as shown in the ANOVA between the items $(F(4, 8176) = 68.2, p<.001)$.

Perceived problems on Twitter are measured with five individual questions, each representing 1) civility of discussions, 2) user banning, 3) content moderation, 4) abuse from other users, and 5) misinformation. Unlike the perceived risk of sharing, the reliability analysis for a scale shows independence among the items (Cronbach's alpha = .61, mean of inter-item correlations = .24). ANOVA between the items $(F(4, 8176) = 679.72, p<.001)$ shows a distinctive difference among the items within the respondents.

The motivation of using Twitter, is measured with a question, "Which would you say is the MOST important reason you use Twitter?" The respondents indicated one from the choices of "Entertainment" (34.2 percent), "A way to stay informed" (31.4 percent), "A way to express my opinions" (6.7 percent), "Keeping me connected to other people" (9.3 percent), "Lets me see different points of view" (10.4 percent), and "It is useful for my job or school" (8 percent). As shown in Figure 1, entertainment and information motivations stood out as the primary reason people use Twitter.

*Figure 1: A Crosstabulation between motivation and regret behavior*

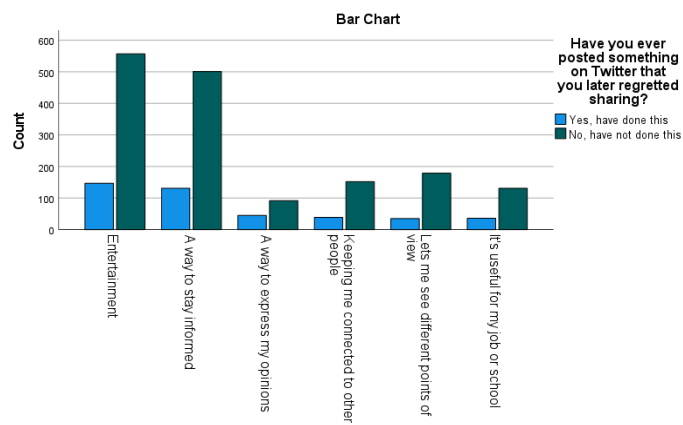

A crosstabulation between motivation (IV3) and regret behavior (DV) shows $\chi(5)$ = 14.4, p<.05, indicating a statistically significant association between motivation and regret behavior; that is, whether Twitter users regret or do not regret sharing on Twitter depend on different motivations of using Twitter.

### 4.3 Inferential statistics

To answer the main research question, logistic regression was performed to ascertain the effects of the perceived risk of sharing, perceived problems on Twitter, and motivation of using Twitter on the likelihood that participants regret sharing on Twitter. The logistic regression model was statistically significant, $\chi2(15) = 102.5$, p < .001. The model explained 7.6 percent (Nagelkerke R squared) of the variance in regret sharing and correctly classified 78.5 percent of cases. Whether Twitter users regret or do not regret sharing on Twitter, depends on different motivations for using Twitter. We observe that "A way to express my opinion" is statistically significant in the model, indicating that the odds of Twitter user regretting sharing for this motivation is 2.1 times higher than the motivation of entertainment. Perceived risks of potential hostility and visibility were negatively associated with an increased likelihood of regret sharing. In contrast, perceived problems on Twitter concerning misinformation were negatively associated with the likelihood of regret sharing.

# 5    Discussion and Conclusion

Researching regret posting on social media has a range of implications. Firstly, it raises questions about the impact of social media on our well-being. Many studies have suggested that using social media may have a negative impact on mental health and psychological wellbeing. Research on regret posting could provide further insights into this issue, considering the psychological consequences of sharing too much personal information online.

Second, the research could affect how people use social media more generally. People who regret posting may be more cautious about sharing information in the future. They may also be more likely to take steps to carefully manage their online profiles and think twice before posting anything that could be seen as impulsive or controversial. As such, the research findings may lead to improved online safety and information management behaviors.

Third, research on regret posting could help inform privacy policies and guidelines designed to protect users. Social media platforms could use the research results to develop better ways to protect users' privacy and combat potential cases of regret posting. They could also use the information to better educate users on the potential risks of sharing personal information online.

Finally, the research could have implications for how people use social media as a form of communication. People may be more aware of their digital footprints and the potentially damaging impact that regret posting can have. As such, people may opt for more measured, controlled forms of communication that are less likely to cause regret or embarrassment.